\documentstyle[mprocl]{article}

\def\be{\begin{equation}}
\def\ee{\end{equation}}
\def\bea{\begin{eqnarray}}
\def\eea{\end{eqnarray}}
\def\lb{\label}

\begin{document}

\title{THE NEAR-HORIZON GEOMETRY OF DILATON-AXION BLACK HOLES}

\author{G. CL\'EMENT}

\address{  
LAPTH (CNRS), B.P.110, 
F-74941 Annecy--le--Vieux cedex, France\\E-mail: gclement@lapp.in2p3.fr}  

\author{D. GAL'TSOV\footnote{Supported by RFBR}}

\address{Department of Theoretical Physics, MSU, 119899, Moscow,
Russia\\
E-mail: galtsov@grg.phys.msu.su}

\maketitle\abstracts{Static black holes of dilaton-axion gravity
become singular in the extreme limit, which prevents a direct
determination of their near-horizon geometry. This is addressed by
first taking the near-horizon limit of extreme rotating NUT-less black
holes, and then going to the static limit. The resulting
four-dimensional geometry may be lifted to a Bertotti-Robinson-like
solution of six-dimensional vacuum gravity, which also gives the 
near-horizon geometry of extreme Kaluza-Klein
black holes in five dimensions.}

The discovery of the AdS/CFT dualities \cite{Ma98} stimulated the
search for geometries containing AdS sectors, which 
typically arise as the near-horizon limit of  BPS black 
holes or p-branes in various dimensions. Here we discuss the 
near-horizon limit of 4-dimensional black holes arising in 
the truncated effective theory of the heterotic string:
dilaton-axion gravity with one Abelian vector field (EMDA). 

The Einstein-frame metrics of the NUT-less rotating extreme black hole 
solutions of EMDA \cite{GK94} are given by
\bea
ds^2 & = & \frac{\Sigma r^2}{\Gamma}\,dt^2 -
\frac{\Gamma}{\Sigma}\sin^2\theta\left( d\varphi - 
\frac{2Ma(r+a)}{\Gamma}dt \right)^2 - 
\Sigma\left( \frac{dr^2}{r^2} + d\theta^2 \right), \\
\Sigma &\! = &\! h - a^2\sin^2\theta, \quad \Gamma = h^2 -
r^2a^2\sin^2\theta, \quad h = r^2 + 2M(r+a), \nonumber 
\eea
$M$ and $a$ being the mass and the rotation parameter.
The horizon $r = 0$ reducing to a point in the static case $a = 0$, we
carry out the near-horizon limit in the rotating case $a \neq
0$. First, we transform to a frame co-rotating with 
the horizon, and rescale time by $t \to (r_0^2/\lambda)t$ ($r_0^2
\equiv 2aM$). Then, we put $r \equiv \lambda x$, $\cos\theta\equiv y$,
and take the limit $\lambda \to 0$, arriving at 
\be\lb{nha}
ds^2 = r_0^2 \left[ (\alpha + \beta y^2) \left( x^2\,dt^2 -
\frac{dx^2}{x^2} - \frac{dy^2}{1 - y^2} \right) -
\frac{1 - y^2}{\alpha + \beta y^2}(d\varphi + x\,dt)^2 \right],
\ee
with $\beta = a/2M, \alpha = 1 - \beta$.
This is similar in form to the extreme Kerr-Newman
near-horizon metric \cite{BH}, both
having the symmetry group $SL(2,R) \times U(1)$, and coinciding in the
extreme Kerr case $M = a$. Now take the static limit $a \to 0$ in 
(\ref{nha}), keeping $r_0^2 = 2aM$ fixed. This yields (for $r_0^2 =
1$) the metric
\be\lb{nh}
ds^2 = x^2\,dt^2 - \frac{dx^2}{x^2} - \frac{dy^2}{1 - y^2} - (1 -
y^2)(d\varphi + x\,dt)^2\,.
\ee
This metric, together with the
associated near-horizon dilaton $\phi$, axion $\kappa$ and gauge potential in
the static limit (independent of the original parameters of the black hole
solution)
\be\lb{bremda}
\phi = 0, \qquad \kappa = -y, \qquad A =
- y(d\varphi 
+ xdt)/\sqrt{2}, 
\ee
constitute a new Bertotti-Robinson-like solution of EMDA, the BREMDA solution.

This solution has properties remarkably similar to those of $AdS_2\times S^2$: 
all timelike geodesics are
confined, and the Klein-Gordon equation is separable in terms of the
usual spherical harmonics. The isometry group $SL(2,R)\times U(1)$,
with $U(1)$ being the remnant of the $SO(3)$ symmetry of $S^2$, is generated
by the Killing vectors $L_{-1}, L_0, L_{1}$ and $L_\phi=\partial_\phi$. 
The first three of these constitute the $sl(2,R)$ subalgebra of an
infinite-dimensional algebra of asymptotic symmetries of
(\ref{nh}), with generators
\be
L_n=t^{-n}\left[\left(t+\frac{n(n-1)}{2x^2t}\right)\partial_t\,+\,x(n-1)\partial_x-
\frac{n(n-1)}{xt}\partial_\phi\right] 
\ee
(where $n \in Z$) satisfying the
Witt algebra $[L_n,L_m]=(n-m)L_{n+m}$ up to terms $O(x^{-4})$. It can be
expected that a representation in terms of asymptotic metric variations
will lead to the  Virasoro extension of this algebra with a classical central
charge, opening the way for microscopic counting of the horizon
microstates of dilaton-axion black holes.

The BREMDA solution does have a close connection with $AdS_2\times S^2$,
which is discovered by lifting it to 6 dimensions. EMDA in 4
dimensions may be shown \cite{6red4} to derive from 6-dimensional
vacuum gravity with 2 commuting spacelike Killing vectors 
$\partial_{\eta}$, $\partial_{\chi}$ via a two-step Kaluza-Klein
reduction together with the assumption of a special relation between
the Kaluza-Klein gauge fields, according to the ansatz  
\bea\lb{an}
& ds_6^2 = ds_4^2 - e^{-\phi}\,\theta^2 - e^{\phi}\,(\zeta +
\kappa\theta)^2,& \\
& \theta = d\chi + A_{\mu}dx^{\mu}, \quad \zeta
= d\eta + B_{\mu}\,dx^{\mu}, \quad
G_{\mu\nu} = e^{-\phi}\tilde{F}_{\mu\nu} - \kappa
F_{\mu\nu} & \nonumber
\eea
($F = dA$, $G = dB$).
Using this ansatz, the BREMDA solution (\ref{nh})-(\ref{bremda}) may be
lifted and rearranged, yielding the vacuum solution BR6 of
6-dimensional gravity 
\be\lb{br6}
ds_6^2 = x^2dt^2  - \frac{dx^2}{x^2} -
\frac{dy^2}{1-y^2} - (1-y^2)d{\varphi}^2 
- (d\chi_+ - \sqrt{2}xdt + \sqrt{2}yd\varphi)^2 -
d\chi_-^2.
\ee
This is the trivial 6-dimensional embedding of a 5-dimensional
metric BR5, which can be shown to be the common near-horizon limit of
all static, NUT-less black holes of 5-dimensional Kaluza-Klein theory.
It enjoys the higher
symmetry group $SL(2,R) \times SO(3) \times U(1) \times U(1)$, but breaks
all fermionic symmetries: no covariantly constant spinors exist.

\eject
\end{document}